# Retraction dynamics of polymer-laden droplets impacting a smooth solid surface

Koji Hasegawa [a,*], Patrick Palmetshofer [b], Anne K. Geppert [c], Bernhard Weigand [b]

[a] *Department of Mechanical Engineering, Kogakuin University, Tokyo 163-8677, Japan*

[b] *Institute of Aerospace Thermodynamics, University of Stuttgart, Pfaffenwaldring 31, Stuttgart, 70569, Germany*

[c] *Leibniz Institute for Materials Engineering, Badgasteiner 3 Bremen, 28359, Germany*

[*] Corresponding author's contact details (K. Hasegawa)
Full postal address: 1-24-2 Nishi-shinjuku, Shinjuku, Tokyo, 163-8677
Telephone: +81-9046254130
E-mail address : kojihasegawa@cc.kogakuin.ac.jp

## ABSTRACT

*Hypothesis:* The presence of a polymer in a droplet plays a significant role in contact line dynamics following droplet impact onto a solid substrate. We hypothesize that the polymer concentration modifies the post-impact wetting behavior.

*Experiments*: In our experiments, we visualize the contact line dynamics of polymer-laden droplets impacting onto a flat sapphire substrate. Using high-speed side and bottom-view imaging, we analyze droplet spreading and retraction over a broad range of Weber numbers (10–700) and polyethylene oxide (PEO) concentrations (0–400 ppm). To systematically characterize the contact line dynamics, we used image processing to quantify the maximum spreading factor, receding velocity, and receding contact angle.

*Findings*: The effect of the polymer on the spreading phase is negligible, as the maximum spreading factor follows classical inertial-capillary scaling, showing that polymer addition produces no measurable change in the maximum spreading under the present conditions. In contrast, the receding phase exhibits a strong dependence on the polymer concentration: increasing the PEO content leads to a significant reduction in the receding velocity and dynamic contact angle, while the spreading dynamics remain water-like. These results are consistent with an additional resistance associated with the receding contact-line region, while the microscopic origin of this resistance remains unresolved. To capture this behavior, we propose a scaling model that relates the normalized retraction velocity to the receding contact angle and impact conditions. The model effectively collapses the experimental data across all investigated concentrations and Weber numbers.

## 1. Introduction

Dynamic wetting at liquid–solid interfaces lies at the heart of numerous natural and technological processes [1,2]. A key challenge in interfacial science is to elucidate how the interplay among capillary, viscous, and interfacial forces governs how liquids wet, spread, and detach from surfaces, thereby controlling the momentum, heat, and mass transfer across interfaces [3]. The impact of a droplet onto a solid surface provides a canonical framework for the study of transient interfacial processes. Upon impact, the inertial force of a droplet drives spreading, whereas the capillary and viscous forces govern the subsequent retraction and relaxation. These fundamental interfacial dynamics play a critical role in various applications, including inkjet printing [4], spray cooling [5], pesticide deposition [6], and surface coating technologies [7]. While the dynamics of Newtonian droplets have been extensively studied, the behavior of viscoelastic or polymer-laden droplets—especially during dynamic phases such as post-impact recoil and retraction—remains less well understood. Even at dilute concentrations, aqueous polymer solutions can exhibit viscoelastic properties [8]. The presence of polymer chains within a droplet significantly influences interfacial dynamics after impact, particularly the motion of the dynamic contact line.

Previous studies have shown that although polymers exert minimal influence during the initial inertial spreading phase, they can dramatically slow down the contact line retraction [6]. Generally, droplet impact outcomes can be categorized into six regimes: deposition, prompt splash, corona splash, receding breakup, partial rebound, and complete rebound [9]. Rebound, in particular, poses a challenge for applications requiring rapid and precise droplet deposition on hydrophobic surfaces. During the relaxation phase, which follows the impact, the elastic energy stored during the droplet deformation drives contact line retraction. If the retraction velocity is sufficient, the droplet may fully rebound from the surface [6].

One strategy to suppress rebound is to alter the surface tension of the liquid by adding surfactants [10]. However, surfactants can introduce complications, such as non-uniform droplet formation and post-impact splashing. As an alternative, Bergeron et al. [6] demonstrated that small amounts of polymer additives could suppress rebound without significantly altering the bulk rheological properties of the fluid. In their study, polyethylene oxide (PEO) was added at 0.01 wt% with a molecular weight of $4 \times 10^6$ g/mol, resulting in a reduced retraction velocity post-impact. They attributed this behavior to elevated extensional viscosity during impact, which was approximately three times greater than the shear viscosity of Newtonian fluids. This increased extensional viscosity dissipated the droplet's kinetic energy more effectively than the shear viscosity, thereby slowing down the retraction.

However, subsequent studies questioned the role of extensional viscosity, noting that it had negligible effect on the spreading phase [11–13]. Both the maximum spreading factors and diameter evolution of polymer solution droplets and Newtonian fluids were found to be similar during spreading, despite the higher extensional viscosity. This discrepancy suggests that the extensional viscosity alone cannot account for the observed behavior [14,15], as viscous dissipation would also be expected to affect the spreading phase.

An alternative explanation was proposed by Bartolo et al. [14], who attributed the reduction in retraction velocity to non-Newtonian normal stresses. They extended lubrication theory by incorporating capillary, shear, and normal stress components to explain the retraction slowdown. In contrast, Bertola [16] raised concerns regarding this interpretation. First, the PEO concentration used in Bartolo's experiments may not have been sufficiently dilute, meaning that increased shear viscosity—not normal stresses—could have been responsible for the observed behavior. Second, Bertola argued that the normal stresses generated near the retracting contact line in dilute solutions were too weak to suppress rebound.

Subsequent studies provided new insights by showing that flexible polymer chains such as λ-DNA or PEO could remain stretched behind the retracting contact line [17,18]. Smith and Bertola [17] proposed that these radially oriented polymer chains create resistance at the contact line, leading to an anti-rebound effect and significantly reduced retraction velocities. Similarly, Smith and Sharp [19] demonstrated that the contact line friction induced by residual polymers increased with retraction velocity, using syringe-driven flows of water and dilute PEO solutions. However, despite this progress, the precise origin of the additional frictional force contributing to rebound suppression remains experimentally unclear [20-22].

Previous studies have therefore established two important features of dilute-polymer droplet impact: the initial spreading remains largely Newtonian-like, whereas contact-line recession can be strongly suppressed [6,12,16,17]. Several mechanisms have been proposed to explain this asymmetry, including enhanced extensional viscosity, non-Newtonian normal stresses, and additional resistance associated with polymer–surface interactions near the receding contact line [14,16–19]. Thus, the qualitative suppression of retraction itself is not unresolved. What remains insufficiently established is how the dynamic wetting state quantitatively relates to the retraction velocity over systematically varied polymer concentration and impact inertia, and how this relationship constrains the competing bulk and contact-line interpretations.

The present study aims to address this gap by focusing on the interactions between polymer chains and solid surfaces. Specifically, we experimentally investigate the spreading and retraction dynamics of PEO-laden droplets upon impact. Using synchronized high-speed imaging captured from multiple viewpoints, the evolution of droplet shape and dynamic contact angle was analyzed across a broad range of Weber and Reynolds numbers. Particular emphasis is placed on the retraction phase, where we examine whether an additional resistance associated with the receding contact-line region contributes to the observed retardation. A scaling model is applied to quantify the retraction dynamics and capture the dependence of the normalized retraction velocity on the dynamic contact angle and impact conditions, providing a predictive framework to describe the dynamic wetting in polymer-laden systems. This work advances our understanding of dynamic wetting in viscoelastic droplets and offers valuable insights for designing fluid systems in droplet-based technologies. The findings lay the foundation for future innovations in applications involving polymer-modified impact dynamics.

## 2. Materials and methods

### 2.1. Experimental design

To investigate droplet impact dynamics, we employed a multi-perspective test rig, previously developed and illustrated in Fig. 1(a) [23]. This experimental setup enables simultaneous recording of the spreading process from four distinct viewpoints. The top and side views are captured using a high-speed camera (FASTCAM SA-X2, Photron Ltd.) via an intermediate image formed near a beam splitter. The same camera records the bottom view, configured in a Total Internal Reflection (TIR) mode, which renders the droplet invisible until it contacts the surface. In this configuration, the contact line appears as a sharp dark boundary, allowing precise tracking of its evolution upon impact. Both top/side and bottom views are synchronized at 20,000 frames per second (fps) with a shutter speed of 11 μs and spatial resolutions of 27.3 μm/pix (top), 15.4 μm/pix (side), and 17.7 μm/pix (bottom). An additional high-speed camera (Chronos 1.4, Kron Technologies) captures a fourth, bird's-eye view at an inclination of approximately 20°, oriented 90° relative to the side view. A flat sapphire substrate was selected to ensure optical transparency for TIR imaging.

Droplets of deionized water (DIN 43530) and PEO solutions were dispensed from a blunt needle (18G, 21G, and 27G, Braun Sterican) connected to a syringe pump (Legato 210, KD Scientific). The droplet contact base diameter and dynamic contact angle were extracted using image processing techniques. All experiments were conducted more than three times at ambient room temperature (~20 °C).

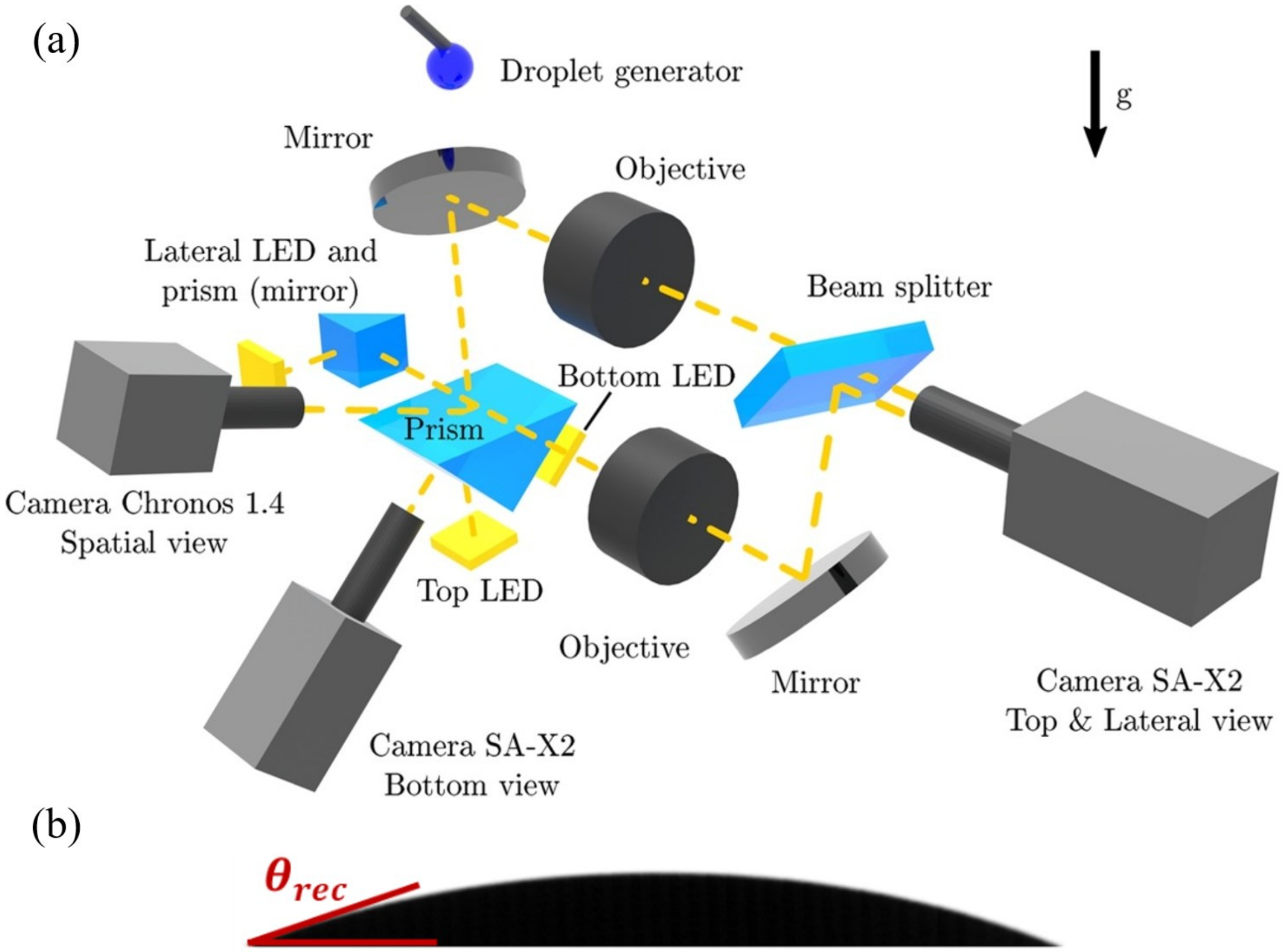

Fig. 1 (a) Schematic of the experimental setup [23]. (b) Representative side-view image illustrating the determination of the apparent dynamic contact angle. The substrate baseline and the locally fitted droplet contour are indicated together with the tangent evaluated at the contact point. The contact angle is defined as the angle between this tangent and the substrate baseline. The values obtained at the left and right contact points were averaged for each frame.

Aqueous solutions of PEO (Product No. 189464, Quality Level 100, Sigma-Aldrich) with a molecular weight $M_w$ of $4\times10^6$ g/mol were used at concentrations of 0 (pure water), 50 (0.005 wt%), 100 (0.01 wt%), 200 (0.02 wt%), and 400 ppm (0.04 wt%). The solutions were prepared using deionized water and thoroughly mixed to ensure homogeneity. The critical overlap concentration $c^*(= 0.77/[\eta])$, estimated using the Mark–Houwink relation $[\eta] = 0.0125 M_w^{0.78}$ [13], was calculated to be 436 ppm. This confirms that even the highest PEO concentration used (400 ppm) remains within the dilute regime. The physical properties of the test samples are summarized in Table 1. Droplets with initial diameters $d_0$ of 2.4 mm, 3.2 mm, and 4.3 mm were released from varying heights $H$ to systematically control the impact velocity $U_0$ (0.61 m/s, 0.95 m/s, 1.34 m/s, 2.30 m/s, and 3.31 m/s), and consequently vary the Weber number $We$ ($= \rho U_0^2 d_0/\sigma$) and Reynolds number $Re$ ($= \rho U_0 d_0/\mu$). Figure 2 illustrates the resulting parameter space for water and 400 ppm PEO solution droplets by plotting in terms of $We$ and $Re$. The dashed line in the figure corresponds to an impact number $P = We/Re^{4/5} = 1$, which distinguishes the dominant energy dissipation mechanism during droplet impact [24]. For $P < 1$ (capillary regime), the droplet's initial kinetic energy is primarily converted into surface energy, whereas for $P > 1$ (viscous regime), viscous dissipation dominates. All tested conditions, across different droplet diameters, impact velocities, and PEO concentrations, fall within the capillary regime. Therefore, surface energy conversion is the dominant dissipation pathway in the present experiments.

Table 1 Thermophysical properties of the test fluids at 20 °C [25].

| Fluids | Concentration $c$ [ppm](wt%) | Density $\rho$ [kg/m$^3$] | Surface tension $\sigma$ [mN/m] | Dynamic viscosity $\mu$ [mPa s] |
|---|---|---|---|---|
| Water | 0 (0) | 1000 | 72 | 1.00 |
| PEO solution | 50 (0.005) | 1000 | 70 | 1.06 |
| | 100 (0.01) | 1000 | 70 | 1.12 |
| | 200 (0.02) | 1000 | 70 | 1.23 |
| | 400 (0.04) | 1000 | 70 | 1.46 |

## 2.2. Image processing and analysis

A checkerboard calibration pattern was used to correct for distortion and to determine the scale and offset for all four viewing perspectives. Impact and spreading parameters were automatically extracted from the bottom, top, and side views using a custom MATLAB script. Following background subtraction and noise removal, the images were binarized for analysis. The final 10 frames prior to impact, captured from the side view, were used to calculate the droplet diameter and impact velocity. Because falling droplets may experience a slight deformation due to aerodynamic forces, the initial diameter $d_0 = \sqrt{4A/\pi}$ was estimated from the average projected area $A$, derived from side-view images [26]. The bottom view enabled precise tracking of the contact line and the measurement of the spreading diameter. The TIR bottom view was used to identify the boundary of the actual wetted area with high contrast. The temporal evolution of the wetted diameter was obtained from this boundary and was used to track the contact-line position and determine the retraction velocity.

The apparent receding contact angle was determined independently from the side-view images using a custom MATLAB routine as shown in Fig. 1(b). Following sub-pixel edge detection, the droplet contour was divided into left and right branches. A substrate baseline, defined for each experimental sequence, was kept fixed throughout the corresponding image sequence. For each branch, contour points near the substrate were selected and fitted with a fourth-order polynomial. The intersection of the fitted polynomial with the baseline defined the contact point, and the local apparent contact angle was calculated from the slope of the fitted polynomial at this intersection relative to the substrate baseline. The left and right contact angles were averaged to obtain the apparent contact angle for each frame. The receding contact angle $\theta_{rec}$ was defined as the temporal average over the quasi-steady plateau during the late retraction stage, typically over approximately 10–30 ms (200–600 frames at 20,000 fps). The automated measurements were independently checked against repeated manual measurements of representative frames; the manual measurements exhibited a standard deviation of approximately 1°.

Measurement deviations in $d_0$, $U_0$, $We$, and $Re$ are summarized in Table 2. The largest possible deviations of $We$ and $Re$ for water droplets were estimated in Table 2 by using the corresponding $U_0$ and $d_0$ = 4.26±0.06 mm as representative values.

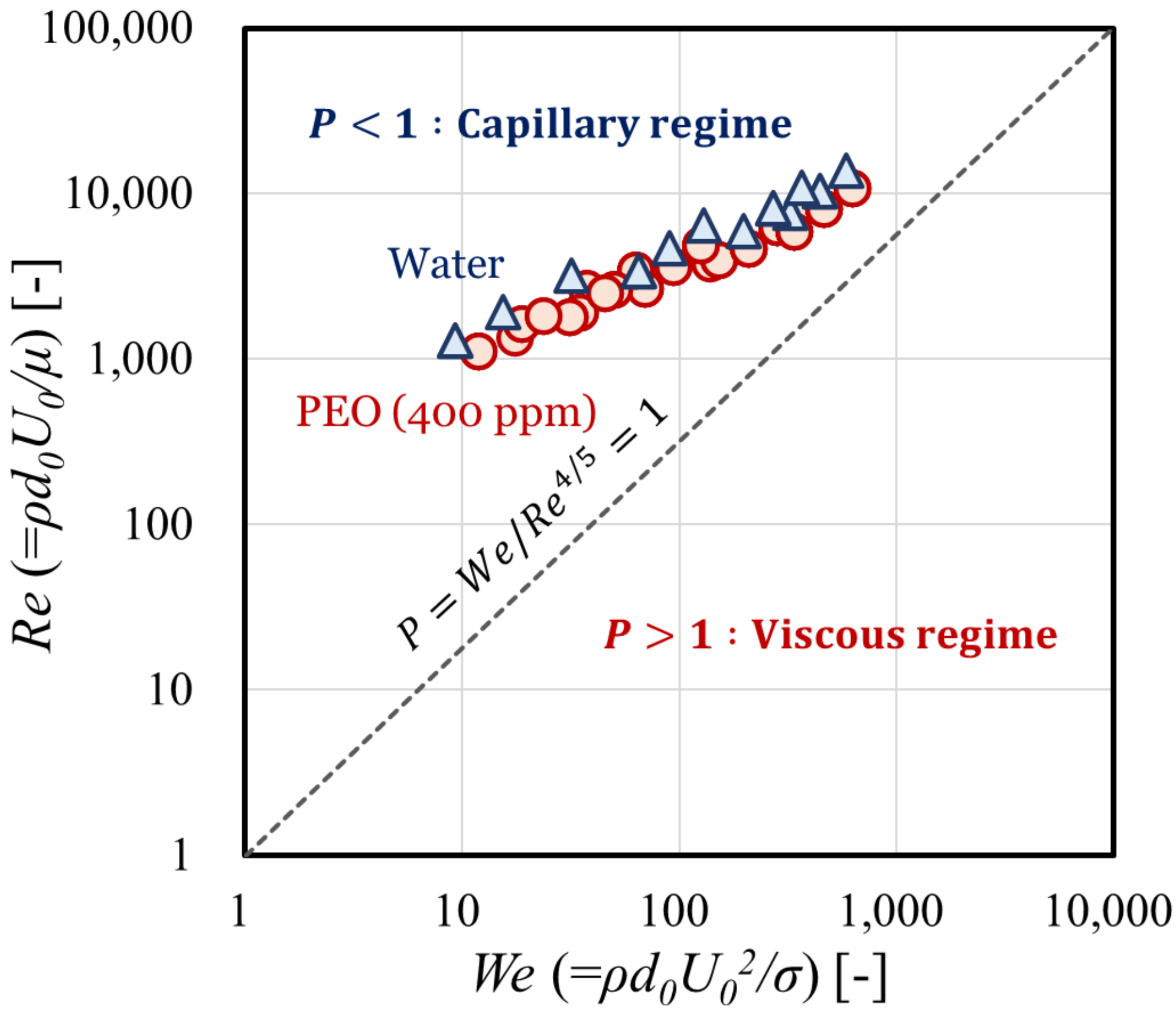

Fig. 2 Parameter space explored in the present study. The dashed line represents the condition $P = 1$, where the impact number $P$ distinguishes between the capillary regime ($P < 1$) and the viscous regime ($P > 1$).

Table 2 Deviations in the measurement of the experimental parameters in this study. Deviations represent the standard deviation from the impact parameters determined from more than three measurements. The initial droplet diameter $d_0$ and impact velocity $U_0$ were controlled by varying the three different needles and five different falling heights, respectively.

| $d_0$ [mm] | 2.36 ± 0.02 | | 3.23 ± 0.04 | | 4.26 ± 0.06 |
|---|---|---|---|---|---|
| $H$ [mm] | 20 | 50 | 100 | 300 | 500 |
| $U_0$ [m/s] | 0.61 ± 0.01 | 0.95 ± 0.01 | 1.34 ± 0.02 | 2.30 ± 0.02 | 3.31 ± 0.06 |
| $We$ [-] | 24 ± 1 | 57 ± 1 | 114 ± 6 | 338 ± 10 | 702 ± 35 |
| $Re$ [-] | 2079 ± 34 | 3211 ± 57 | 4534 ± 149 | 7805 ± 167 | 11248 ± 363 |

### 2.3. Spreading and retraction dynamics: Scaling approach

To analyze the droplet deformation and motion following impact, we employed scaling relations for both the spreading and retraction phases. These relations are derived from force-balance principles and standard assumptions governing inertial-capillary and interfacial dynamics.

During the spreading phase, the maximum spreading diameter—normalized by the initial droplet diameter—is predominantly determined by the balance between inertial and capillary forces, assuming viscous effects are negligible. Accordingly, the scaling law is expressed as:

$$\frac{d_{max}}{d_0} \sim We^{1/4} \tag{1}$$

This well-established scaling law has been extensively validated for Newtonian fluids [24] and is used in the present study as a reference for assessing the behavior of polymer solution systems.

During the retraction phase, the droplet retracts from its maximum spread radius primarily due to surface tension forces. In Newtonian fluids, the retraction velocity is typically described by the Taylor–Culick velocity [27] or capillary–inertial scaling models [28]. Previous studies have proposed that dilute polymers can introduce an additional resistance near the moving contact line, potentially involving polymer–surface interactions and chain deformation [19]. To examine whether the experimentally measured wetting state captures the resulting retraction dynamics, we employ a dimensionless scaling framework that incorporates the receding contact angle. The normalized retraction velocity is defined as follows [28]:

$$\frac{U_{rec}/R_{max}}{U_0/R_0} \sim \left(\frac{1-\cos\theta_{rec}}{We}\right)^{1/2}, \tag{2}$$

where $U_{rec}$ denotes the retraction velocity, $R_{max}$ is the maximum radius attained at full droplet spreading, $R_0$ is the initial droplet radius, and $\theta_{rec}$ represents the receding contact angle. Equation (2) enables comparative analysis across varying polymer concentrations and impact conditions by relying on experimentally accessible parameters. This theoretical framework will be applied in Section 3.3 to validate the experimental data and to elucidate the retraction dynamics observed in polymer-laden systems.

## 3. RESULTS AND DISCUSSION

### 3.1. Effect of polymer additive

The impact dynamics of water and aqueous polymer solutions were recorded using a high-speed camera, as illustrated in Fig. 3. Figures 3(a) and 3(b) show that both water and polymer droplets spread and retract smoothly after reaching their maximum spreading diameter ($t/\tau = 2.4$). However, their receding behaviors differ notably: the water droplet retracts faster than the PEO solution droplet. Comparing Figs. 3(b) and 3(c) reveals that the droplet interface becomes increasingly wavy as the Weber number rises from 100 to 500, indicating enhanced interfacial instability at higher $We$ values. Furthermore, this wavy interface persists into the later stages of the receding phase ($t/\tau = 38.5$).

To elucidate the effect of polymer additives, Fig. 4(a) presents the time evolution of the droplet diameter derived from Figs. 3(a) and 3(b). While the spreading phases for both water and PEO droplets collapse onto a single curve, a pronounced difference emerges during retraction: the water droplet retracts rapidly after reaching its maximum diameter, whereas the polymer droplet recedes more slowly. The central question addressed in this study is the underlying cause of this reduction in receding velocity, which is discussed in Section 3.2. Prior to focusing on retraction, we quantitatively analyze the spreading phase of both fluid types.

Figure 4(b) illustrates the influence of the Weber number on the maximum spreading factor $d_{max}/d_0$, as indicated by the leading lines. As expected, $d_{max}/d_0$ increases with $We$. The data further show that the spreading behavior overlaps across all cases, whereas the slope of $d/d_0$ during the receding phase varies systematically with the impact condition. Thus, although the post-maximum-spreading retraction is primarily governed by capillary–inertial dynamics, its dimensional velocity retains a moderate dependence on the preceding impact conditions. This dependence is examined quantitatively below in relation to the maximum spreading geometry and the receding contact angle.

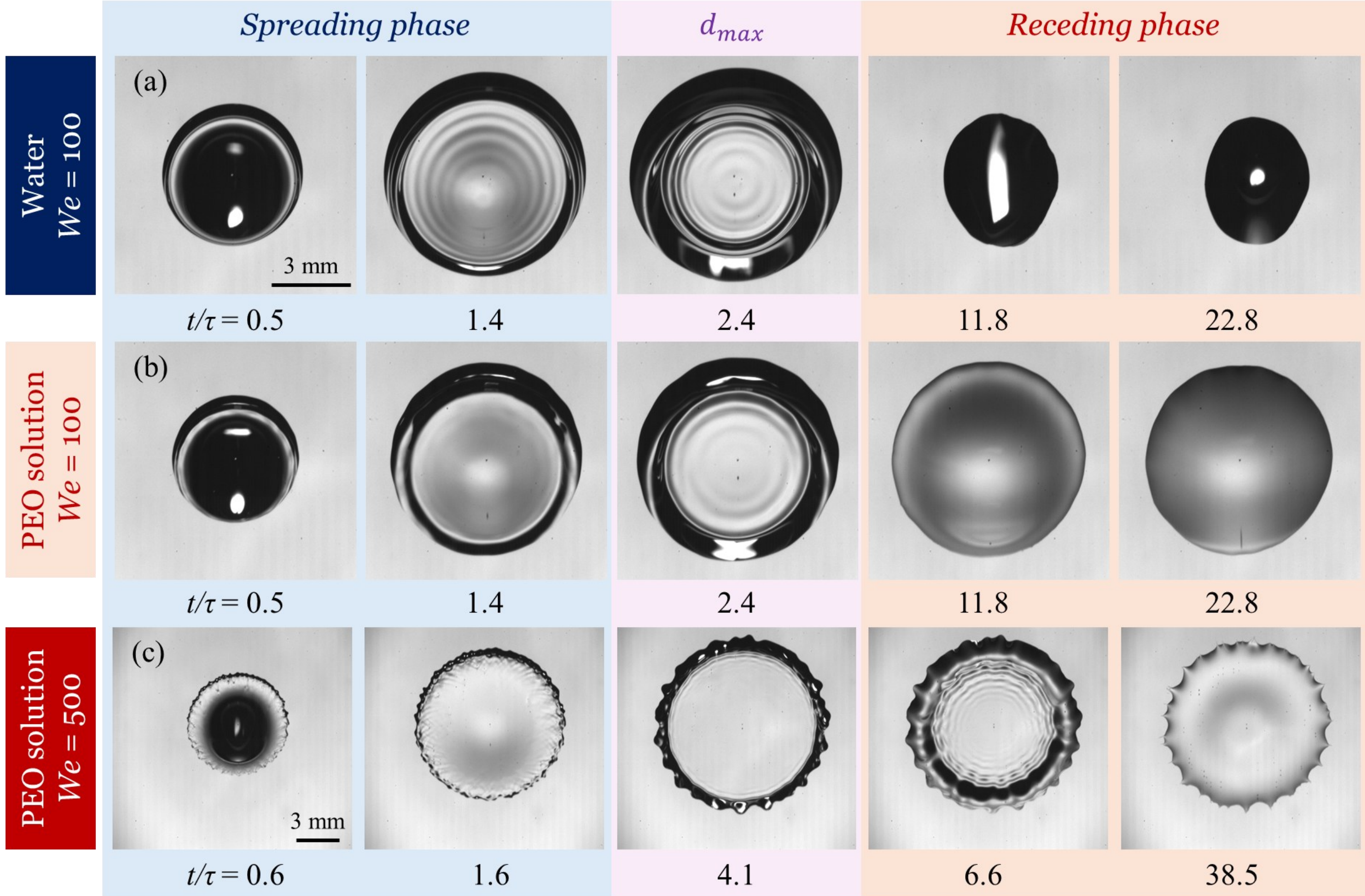

Fig. 3 Snapshots illustrating the spreading and retraction phases following droplet impact for (a) a water droplet at $We = 100$, (b) a PEO solution droplet at 400 ppm concentration with $We = 100$, and (c) a PEO solution droplet at $We = 500$. Time $t$ is normalized by the impact time scale $\tau = (d_0/U_0)$. See also the supplementary material.

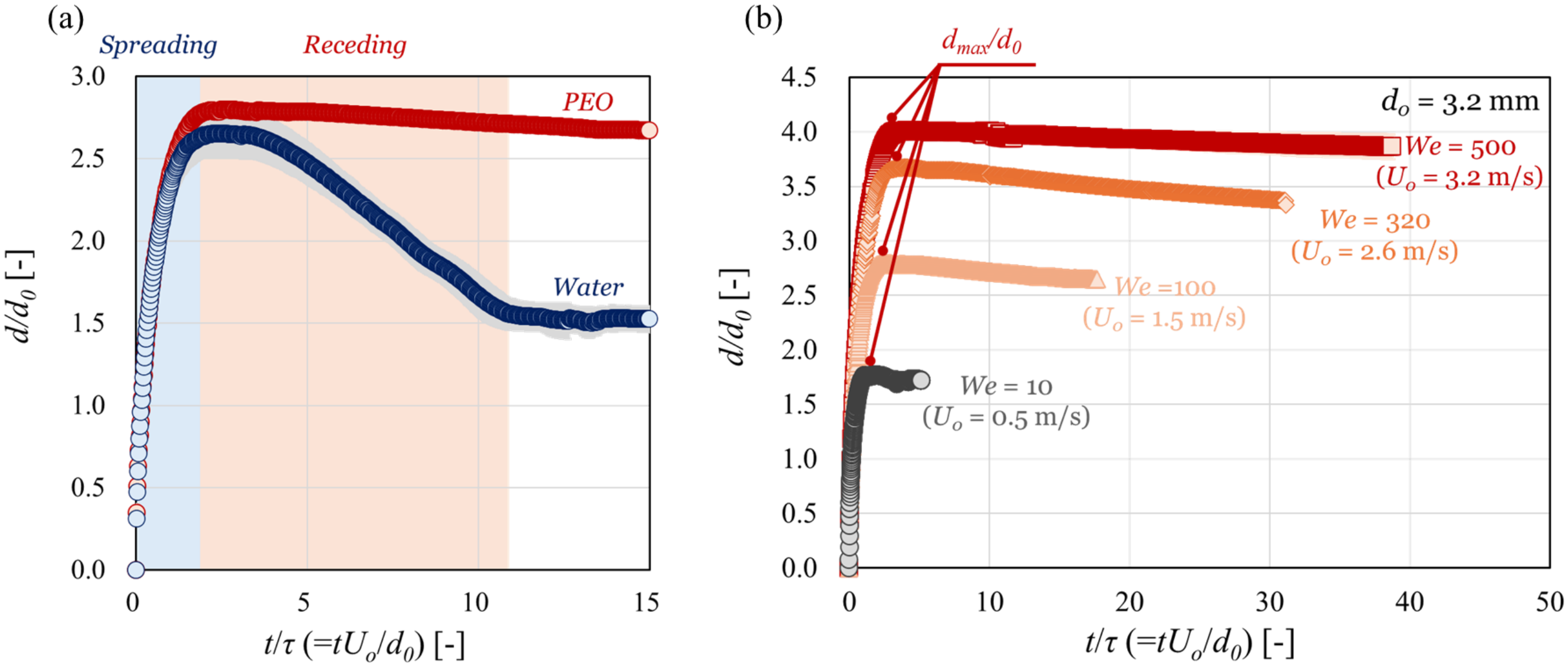

Fig. 4 Nondimensional time evolution of normalized droplet diameter during spreading and retraction: (a) comparison between water and PEO solution droplets at $We = 100$, and (b) effect of Weber number on the impact behavior of PEO droplets with $d_0$ =3.2 mm and varying $U_0$ as indicated. Error bars represent the standard deviation for three measurements.

Figure 5 presents the systematic effect of the Weber number on the normalized maximum spreading factor $d_{max}/d_0$. The experimental data show no significant difference between water droplets and PEO solution droplets, indicating that inertial forces dominate over the rheological effects of polymer additives in this regime. Importantly, the results exhibit strong quantitative agreement with the predictions of the scaling law given by Eq. (1), consistent with findings from previous studies [13]. This agreement further validates the applicability of the scaling relation expressed in Eq. (1) for describing droplet spreading dynamics under the tested conditions.

As noted, the spreading phase of both water and polymer droplets can be accurately described by the classical scaling law in Eq. (1). However, the droplet dynamics during the receding phase remain unclear and require further investigation. To quantitatively analyze the retraction behavior, the receding velocity was extracted from experimental data.

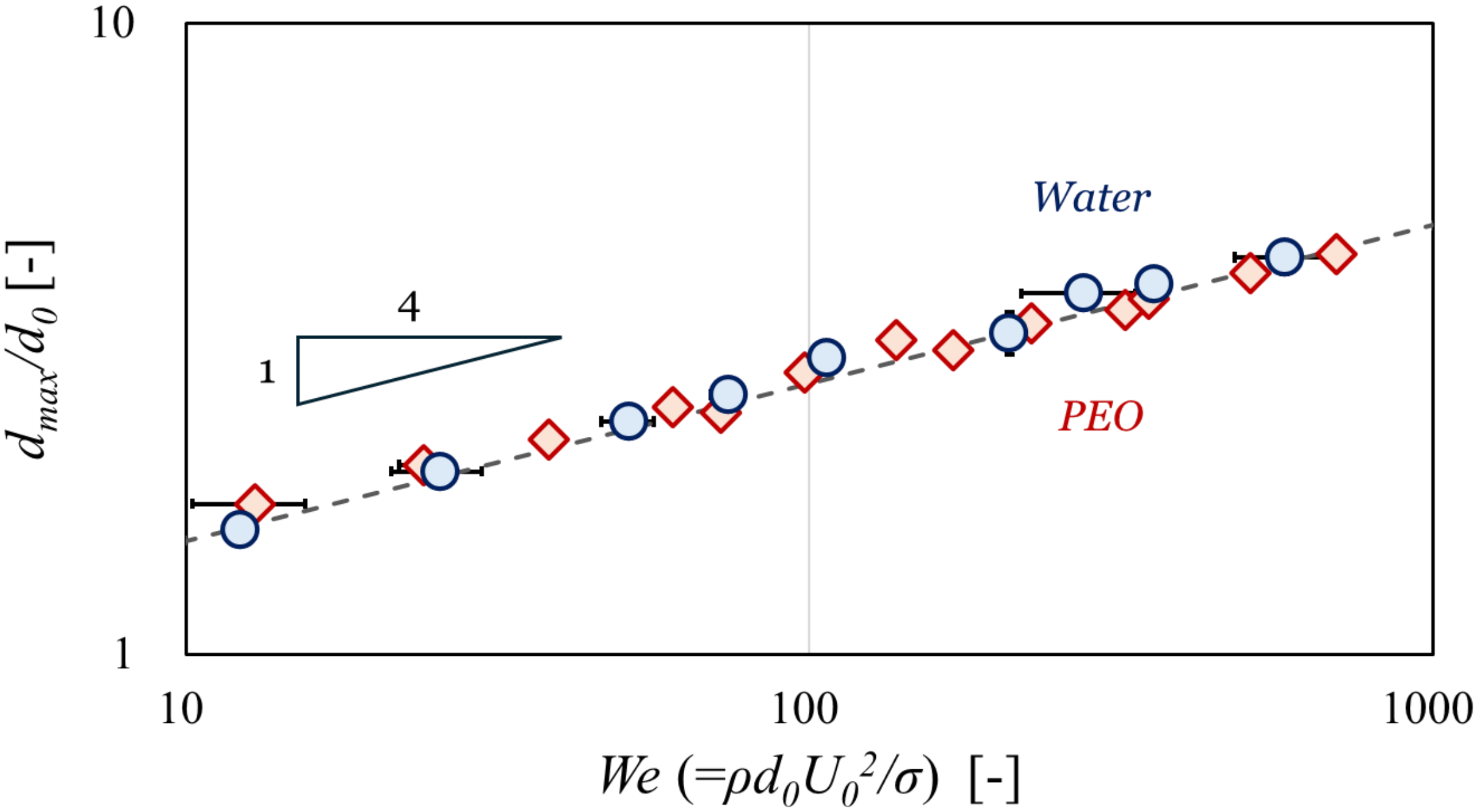

Fig. 5 Maximum spreading factor $d_{max}/d_0$ for water and PEO solution droplets across a wide range of Weber numbers. The dashed line represents the scaling law given by Eq. (1). Error bars represent the standard deviation for three measurements.

Figure 6(a) presents the retraction velocity of both water and PEO solution droplets over a broad range of Weber numbers. Water droplets exhibited relatively constant retraction velocities, averaging approximately 200 mm/s. In contrast, the retraction velocities of PEO solution droplets were one to two orders of magnitude lower. The droplet interface during retraction offers important insights into this discrepancy.

Figure 6(b) shows that the receding PEO droplet develops a pronounced azimuthally corrugated contact-line morphology, whereas the water droplet in Fig. 6(c) retains a comparatively smooth wetted boundary. This pattern is observed after maximum spreading during the retraction stage and therefore does not correspond to a corona splash, which is associated with the lifting of the expanding lamella during the earlier spreading stage. The TIR configuration directly visualizes the wetted region on the sapphire substrate; accordingly, the structures in Fig. 6(b) represent irregularities of the receding contact-line boundary. We note that such an irregular morphology is not considered, by itself, evidence of a polymer-specific microscopic mechanism [29].

To examine whether the polymer-containing droplet leaves persistent material on the substrate, the impacted droplets were further observed after complete evaporation. As shown in Fig. 6(d), the 400 ppm PEO droplet left a spatially extended residual footprint over the previously wetted region, including peripheral and filament-like deposits. In contrast, the water control in Fig. 6(e) showed only a small localized residue and no comparable extended footprint. These post-drying observations are consistent with persistent polymer–substrate interactions. However, because polymer redistribution may also occur during the subsequent evaporation process, they do not establish whether adsorption occurs during the millisecond retraction stage. The following subsection discusses possible physical origins of the observed retraction resistance.

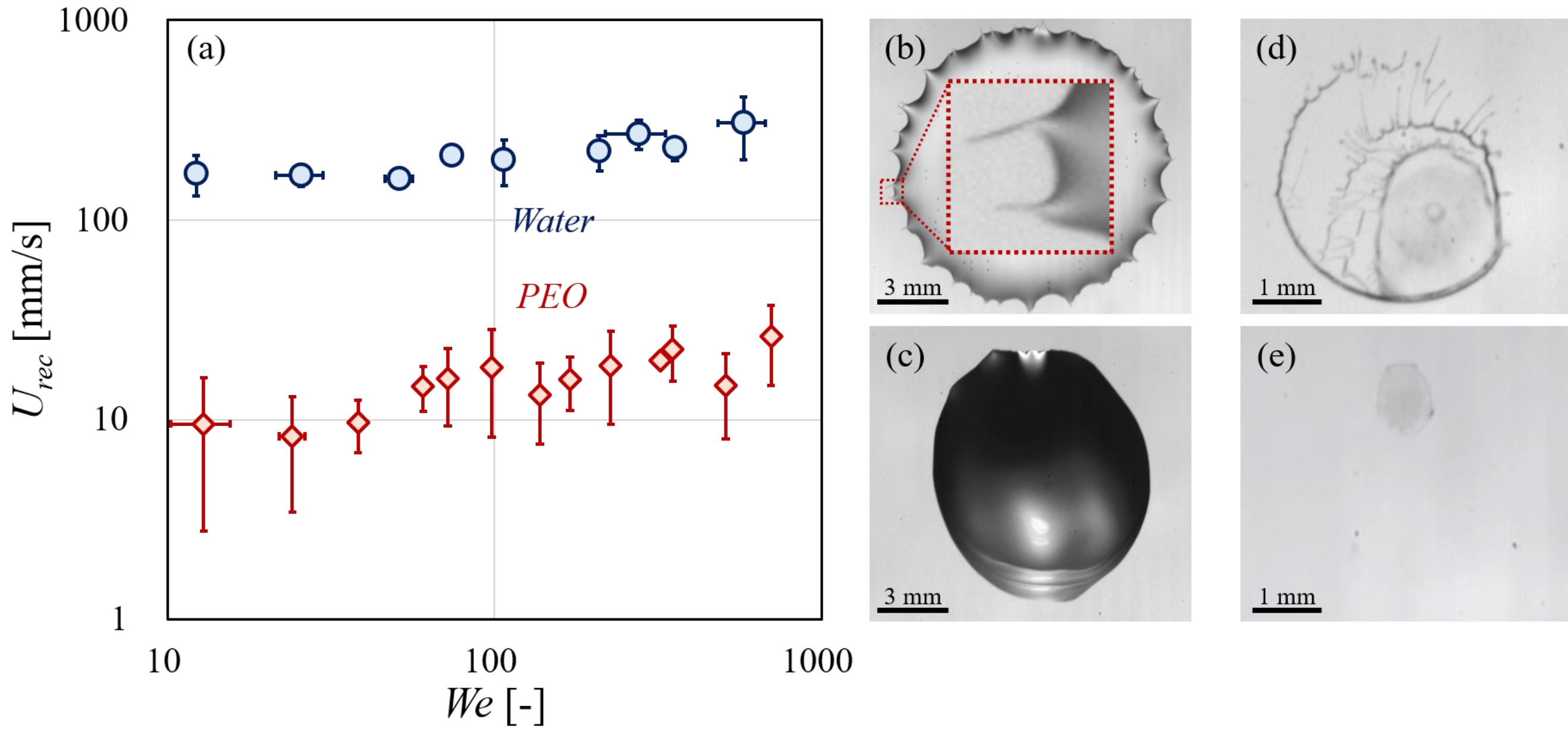

Fig. 6 Retraction dynamics of water and PEO solution droplets: (a) retraction velocity across various Weber numbers. Error bars represent the standard deviation for three measurements. Representative snapshots represent droplets retraction for (b) PEO solution and (c) water droplets at $We = 500$, and footprints of (d) PEO solution and (e) water droplets on the solid surface after complete evaporation of droplets.

### 3.2. Effect of polymer concentration

Varying the polymer concentration provides an effective approach to investigate polymer-induced resistance at the contact line. Figure 7(a) presents the retraction velocity measured by systematically varying the Weber number and the polymer concentration. At each PEO concentration, the dimensional retraction velocity exhibits a moderate but systematic increase with $We$, with the dependence becoming more evident at higher polymer concentrations. This trend is weaker than the pronounced concentration dependence: at comparable $We$, increasing the PEO concentration markedly reduces $U_{\mathrm{rec}}$. The observed $We$ dependence is consistent with the capillary–inertial framework of Eq. (2). Combining Eq. (2) with the experimentally observed $R_{\mathrm{max}}/R_0 \sim We^{1/4}$ scaling shows that, for a receding contact angle that varies only weakly with $We$, the dimensional retraction velocity is expected to increase moderately with impact inertia.

The corresponding receding capillary number, $Ca_{\mathrm{rec}} = \mu U_{\mathrm{rec}}/\sigma$, decreases systematically with increasing PEO concentration, reflecting the strong suppression of contact-line motion despite the modest increase in shear viscosity. For reference, the water droplets with $U_{\mathrm{rec}} \approx 0.2$ m/s correspond to $Ca_{\mathrm{rec}} \approx 2.8 \times 10^{-3}$, whereas the highest PEO concentration yields values of order $10^{-4}$–$10^{-3}$.

Figure 7(b) summarizes the effect of PEO concentration by averaging the retraction velocities from Fig. 7(a). The retraction velocity decreases nonlinearly with increasing polymer concentration. Even low PEO concentrations (e.g., 50 ppm) significantly reduce the retraction velocity, with the reduction becoming more pronounced at higher concentrations. The pronounced reduction in retraction velocity cannot be explained by the modest increase in shear viscosity alone. Across the present conditions, the viscosity increases from 1.00 mPa s for water to 1.46 mPa s for the 400 ppm PEO solution, while the corresponding Ohnesorge numbers $Oh\left(= \mu/\sqrt{\rho\sigma d_0}\right)$ remain between approximately $1.8 \times 10^{-3}$ and $3.6 \times 10^{-3}$. Nevertheless, a low Ohnesorge number does not exclude elastic or extensional effects in dilute polymer solutions. Using the empirical correlation originally reported by Kalashnikov and Askarov [30] for dilute aqueous PEO solutions, the characteristic relaxation time for the present PEO with $M_w = 4 \times 10^6$g/mol is estimated to increase from approximately 15 ms at 50 ppm to 43 ms at 400 ppm at 20 °C. These values are comparable to the capillary–inertial timescale of the present droplets (~ 14–33 ms), indicating that elastic memory may persist during the post-impact dynamics. Thus, the present results do not exclude a contribution from bulk viscoelastic stresses. Expressed as a capillary–inertial Deborah number, $De_{ci} = \tau/t_{ci}$, the present PEO conditions correspond to $De_{ci} \approx 0.46$–$3.2$. The resulting $De_{ci} = O(1)$ confirms that elastic memory may coexist with the observed contact-line-associated resistance. However, previous experiments indicate that bulk viscoelasticity alone is insufficient to account for the pronounced suppression of retraction. When the interaction with an extended solid substrate was strongly reduced in small-target impacts, polymer additives produced little change in lamella retraction [12]. Similarly, experiments under Leidenfrost conditions showed a strongly diminished polymer effect when direct liquid–solid wetting was suppressed [16]. Moreover, particle-velocimetry measurements showed comparable bulk velocity fields for water and dilute PEO droplets during both spreading and retraction, whereas the contact-line velocity of the polymer solution was approximately one order of magnitude lower [31]. These observations point to an additional resistance associated with the receding contact-line region, while not excluding concurrent bulk viscoelastic contributions.

To examine whether the observed concentration dependence is consistent with the contact-line force picture proposed in Ref. [18], we adopt its order-of-magnitude polymer-force scaling,

$$F_p \sim \sqrt{n} f_p = \sqrt{n} \frac{k_B T L}{R_o^2} \mathcal{L}^{-1}(l) \tag{3}$$

The bulk number density of polymer coils at the droplet's contact line is given by $n = \frac{\rho_p c_0 N_A}{M_w}$, where $\rho_p$ is the polymer density, $N_A$ is Avogadro's number, $c_0$ is the polymer's volume concentration, and $M_w$ is its molecular weight. The frictional force exerted by a single polymer chain, $f_p$, is expressed as $f_p(= \frac{k_B T}{R_o^2} L\mathcal{L}^{-1})$, where $k_B$ is the Boltzmann constant, $T$ is the absolute temperature, $R_o$ is a characteristic polymer length scale, and $\mathcal{L}^{-1}(l)$ is the inverse Langevin function, which can be approximated as [18]:

$$\mathcal{L}^{-1}(l) = \frac{3l - (l/5)(6l^2 + l^4 - 2l^6)}{1 - l^2} \tag{4}$$

The polymer chain is modeled as a single spring with fractional extension $l = r/L$, where $r$ is the polymer elongation and $L$ is the fully stretched chain length. According to a previous study [18], the fractional stretching of polymer molecules is approximately 50%, based on direct observations of individual polymer conformational dynamics in steady shear flow. Correspondingly, the inverse Langevin function evaluates to $\mathcal{L}^{-1}(0.5) = 1.8$. Additionally, polymer coils stretched by the contact line as it sweeps the substrate align within a thin interfacial layer, causing the polymer force $F_p$ to scale as $\sqrt{n}$, where $n$ is the bulk number density. The predicted $F_p$ based on this microscopic polymer model is presented in the inset of Fig. 7(b). Because the main panel and inset represent different physical quantities, no direct numerical correspondence between their ordinates is implied. Rather, the inset illustrates the modelled concentration dependence of the polymer-associated resistance: the estimated $F_p$ increases with increasing PEO concentration, whereas the experimentally measured $U_{\mathrm{rec}}$ decreases. This inverse correspondence is qualitatively consistent with an increasing resistance to contact-line recession. However, the agreement in concentration trend alone neither constitutes quantitative validation of the limiting microscopic model nor establishes polymer adsorption as the unique origin of the resistance.

Even though we assumed the phenomenological concentration scaling by $\sqrt{n}$ in Eq. (3), the concentration dependence predicted by this model is qualitatively consistent with the observed increase in retraction resistance. However, this agreement alone does not establish polymer adsorption as the microscopic origin of the resistance. Although the microscopic origin of the additional resistance remains unresolved, its macroscopic manifestation can be examined through the receding wetting state. In particular, the receding contact angle provides a direct measure of the capillary driving condition associated with contact-line motion. Therefore, we analyze the physical relationship between retraction velocity and receding contact angle using both experimental data and theoretical predictions.

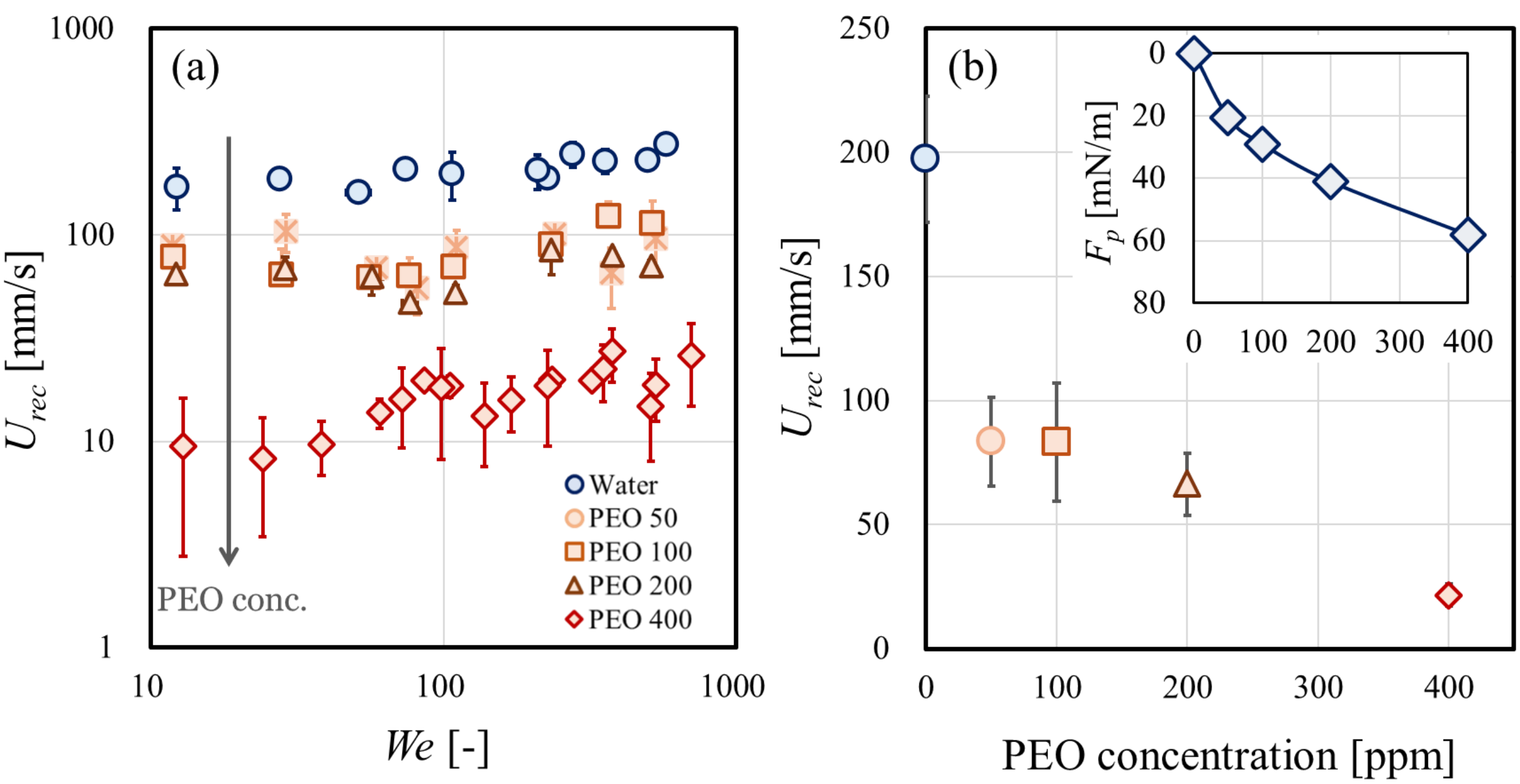

Fig. 7 Effect of polymer concentration on retraction dynamics for $c = 0$ (water), 50, 100, 200, and 400 ppm: (a) retraction velocity across varying $We$; (b) retraction velocity as a function of polymer concentration. The inset shows the order-of-magnitude polymer-associated line force estimated using Eq. (3). Error bars represent the standard deviation for three measurements.

### 3.3. Model validation via experimental data

Figure 8(a) illustrates the effect of the Weber number on the normalized retraction dynamics. The normalized retraction velocity is defined as the ratio of the retraction rate $U_{rec}/R_{max}$ after reaching the maximum spreading diameter to the initial shear rate $U_0/R_0$ imposed by the droplet impact, as expressed in Eq. (2). The model predictions, shown as dashed lines, closely match the experimental data across different PEO concentrations, corresponding to the measured receding contact angles. Notably, while the receding contact angle is significantly influenced by PEO concentration, it remains nearly independent of the Weber number, consistent with observations for Newtonian systems [32]. The concentration dependence of $\theta_{rec}$ should be distinguished from the nearly concentration-independent maximum spreading observed during the earlier inertial stage. The maximum spreading factor primarily reflects the inertia–capillary balance immediately after impact, whereas $\theta_{rec}$ characterizes the dynamic wetting state during the subsequent contact-line recession. The systematic decrease in $\theta_{rec}$ with increasing PEO concentration therefore indicates an increasing resistance to contact-line recession rather than a modification of the initial inertial spreading. The microscopic origin of this resistance cannot be uniquely identified from the contact-angle measurement alone.

To assess the generality of the retraction scaling law by Eq. (2), Fig. 8(b) compares the experimental results with the model predictions, demonstrating a clear data collapse. The collapse indicates that the measured receding contact angle provides an effective macroscopic descriptor of the retraction dynamics across the investigated polymer concentrations and impact conditions. Importantly, this agreement should not be interpreted as direct evidence for a specific microscopic dissipation mechanism. Rather, it demonstrates that the polymer-induced retardation is systematically coupled to the receding wetting state, consistent with an additional resistance associated with the moving contact-line region. Determining the molecular origin of this resistance, including the possible roles of polymer adsorption, chain deformation, and local concentration redistribution, requires direct time-resolved measurements beyond the scope of the present study.

While the model expressed by Eq. (2) provides an accurate framework for analyzing the retraction characteristics in this study, further validation is required to extend its applicability to higher polymer concentrations, different polymer types, and various substrate materials. Additionally, direct visualization of polymer distribution within the droplet and adsorption dynamics during impact, spreading, and retraction remains critical. These topics are beyond the scope of the present work but will be explored in future research.

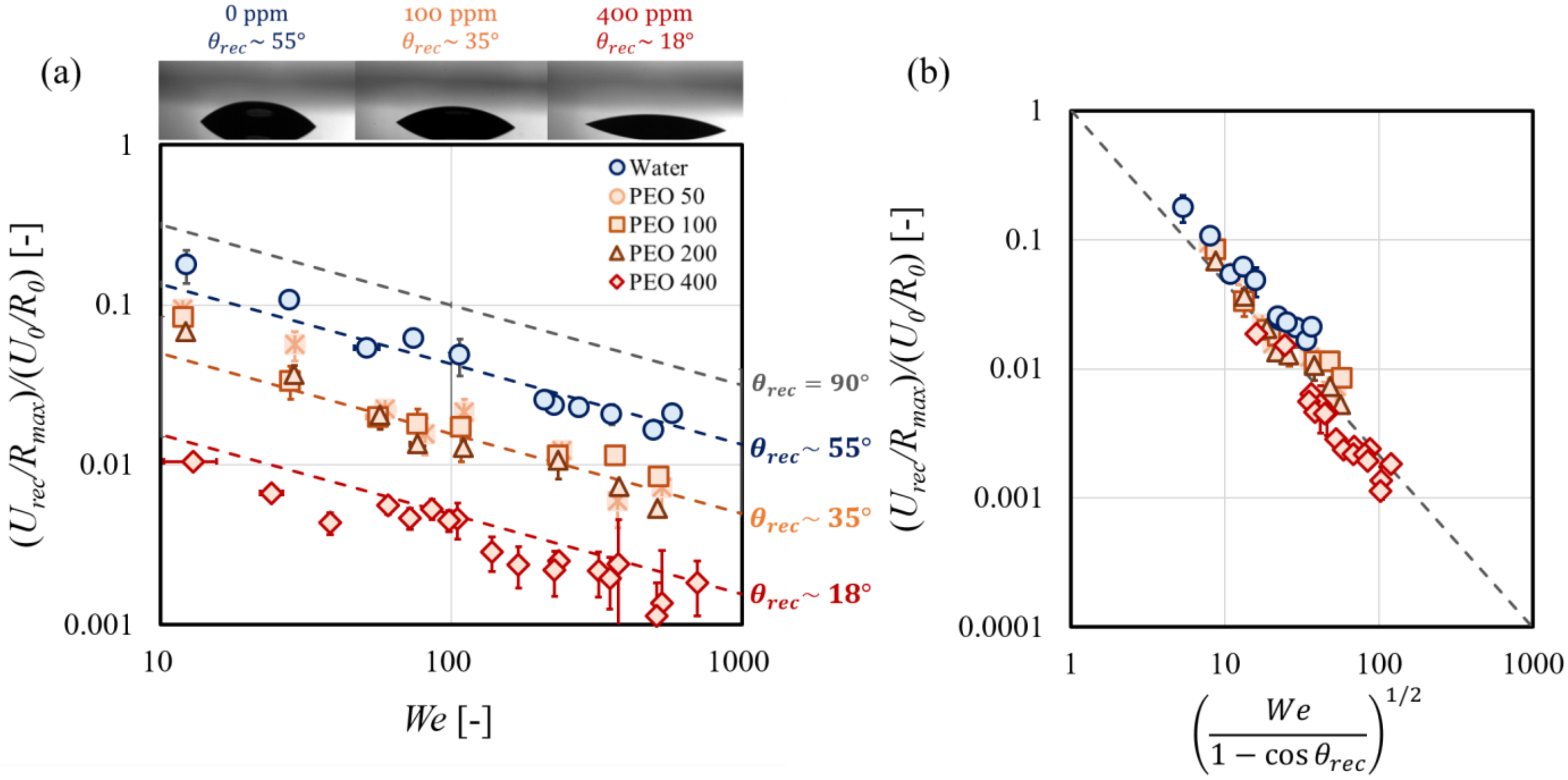

Fig. 8 Model validation with systematic experimental data: (a) normalized retraction characteristics for varying $We$. Insets show receding droplet shapes for $c = 0$ (water), 100, and 400 ppm with corresponding measured receding contact angles. Dashed lines represent model predictions from Eq. (2) for $\theta_{rec} \sim 18°$, 35°, 55°, and 90° by Eq. (2). (b) data collapse for all investigated experimental cases, with the dashed line indicating the model prediction. Error bars represent the standard deviation for three measurements.

## 4. CONCLUSIONS

This study conducted a comprehensive and systematic experimental investigation of the impact, spreading, and receding dynamics of water (0 ppm) and PEO solution droplets (50, 100, 200, and 400 ppm) on a flat sapphire substrate across a broad range of Weber numbers (10–700). Our findings confirm that the initial spreading behavior is predominantly governed by classical inertial–capillary scaling and remains unaffected by polymer concentrations up to 400 ppm. This observation is consistent with recent theoretical and numerical insights suggesting that viscous boundary layers dominate the early stages of high-speed impact regardless of bulk viscoelastic properties [33, 34]. The modest increase in shear viscosity and the small Ohnesorge numbers ($Oh = O(10^{-3})$) indicate that ordinary shear-viscous dissipation alone cannot account for the pronounced retraction retardation. At the same time, the estimated characteristic polymer relaxation times are comparable to the post-impact capillary–inertial timescale, indicating that bulk elastic effects may persist during retraction and therefore cannot be excluded.

The central finding of the present study is that the strong reduction in retraction velocity is systematically accompanied by a decrease in the receding contact angle. Together with previous observations showing that polymer-induced retardation is strongly diminished when liquid–solid interaction is suppressed [12,16] and that bulk velocity fields remain similar while contact-line motion is strongly retarded [31], the present results support an additional resistance associated with the moving contact-line region rather than a mechanism governed solely by bulk viscoelasticity.

To quantitatively capture these complex dynamics, we successfully validated a predictive scaling framework that relates the normalized retraction velocity to the dynamic receding contact angle and impact conditions. By utilizing synchronized high-speed imaging captured from multiple viewpoints, including a bottom view configured in a TIR mode, we achieved precise tracking of the contact line evolution on the sapphire substrate. Incorporating the experimentally measured receding contact angle into the retraction scaling organizes the data across the investigated PEO concentrations and impact conditions onto a common relation. This collapse establishes a quantitative connection between the macroscopic retraction dynamics and the receding wetting state, but does not uniquely identify the microscopic origin of the additional resistance. The persistent residual footprint observed after complete evaporation of PEO droplets is consistent with polymer–substrate interactions; however, it does not establish adsorption during the millisecond retraction stage. Polymer adsorption, chain deformation, and local concentration redistribution near the receding meniscus therefore remain plausible microscopic contributions that require direct time-resolved verification. Our experimental findings further reinforce the numerical simulation results regarding three-phase contact line dynamics with polymer additives [20].

These findings provide a quantitative framework for relating polymer-modified retraction to the receding wetting state, with potential relevance to droplet-deposition processes. Future work should focus on time-resolved, molecular-scale measurements of polymer distribution and surface association during the millisecond retraction stage and on testing the present framework across different polymer–substrate combinations.

## CRediT authorship contribution statement

**Koji Hasegawa:** Conceptualization, Methodology, Investigation, Validation, Software, Formal analysis, Data curation, Visualization, Funding acquisition, Writing – original draft, Writing – review & editing. **Patrick Palmetshofer:** Methodology, Software, Writing – review & editing. **Anne K. Geppert:** Conceptualization, Resources, Writing – review & editing. **Bernhard Weigand:** Conceptualization, Supervision, Project administration, Funding acquisition, Writing – review & editing,

## Declaration of competing Interest

The authors declare that they have no known competing financial interests or personal relationships that could have appeared to influence the work reported in this paper.

## Acknowledgements

This work was supported by JSPS KAKENHI Grant Number 23KK0261. We appreciate the financial support of this work by the Deutsche Forschungsgemeinschaft (DFG) in the framework of the International Research Training Group “Droplet Interaction Technologies” (GRK 2160/2: DROPIT) under project number 270852890

## Appendix A. Supplementary material

Supplementary material related to this article can be found online at (URL).

## Data availability

Data will be made available on request.